\documentstyle[12pt]{article}
\input psfig.tex
\begin{document}

\def\ov{\overline}
\def\ra{\rightarrow}
\def\pslash{\not{\hbox{\kern -1.5pt $p$}}}
\def\kslash{\not{\hbox{\kern -1.5pt $k$}}}
\def\aslash{\not{\hbox{\kern -1.5pt $a$}}}
\def\bslash{\not{\hbox{\kern -1.5pt $b$}}}
\def\Dslash{\not{\hbox{\kern -4pt $D$}}}
\def\wslash{\not{\hbox{\kern -4pt $\cal W$}}}
\def\zslash{\not{\hbox{\kern -4pt $\cal Z$}}}
\def\kln{\kappa_{L}^{NC}}
\def\krn{\kappa_{R}^{NC}}
\def\klc{\kappa_{L}^{CC}}
\def\krc{\kappa_{R}^{CC}}
\def\bbz{{\mbox {\,$b$-${b}$-$Z$}\,}}
\def\ttz{{\mbox {\,$t$-${t}$-$Z$}\,}}
\def\ttzz{{\mbox {\,$t$-${t}$-$Z$-$Z$}\,}}
\def\ttww{{\mbox {\,$t$-${t}$-$W$-$W$}\,}}
\def\wwtt{{\mbox {\,$W^+W^-t{\bar t} \;$}\,}}
\def\tta{{\mbox {\,$t$-${t}$-$A$}\,}}
\def\bba{{\mbox {\,$b$-${b}$-$A$}\,}}
\def\tbw{{\mbox {\,$t$-${b}$-$W$}\,}}
\def\Tr{{\rm Tr}}

\setcounter{footnote}{1}
\renewcommand{\thefootnote}{\fnsymbol{footnote}}

\begin{titlepage}

{\small
\noindent
{hep-ph/9709316} \hfill {CINVESTAV/FIS-51/97}  \\
{ September 1997} \hfill {ANL-HEP-PR-97-63}\\
\rightline{ {MSUHEP-70815} } 
}

\vspace{2.0cm}

\centerline{\Large\bf Anomalous \wwtt ~couplings at}
\centerline{\Large\bf the $e^+ e^-$ Linear Collider}

\vspace*{1.2cm}
\baselineskip=17pt
\centerline{\normalsize  
F. Larios,\footnote{
\baselineskip=12pt  Also at the Departamento de F\'{\i}sica,
CINVESTAV, Apdo. Postal 14-740, 07000 M\'exico, D.F., M\'exico.}
Tim Tait\footnote{
\baselineskip=12pt  Also at Argonne National Laboratory,
HEP Division, 9700 South Cass Avenue,  Argonne, IL 60439.},
  and C.-P. Yuan}

\centerline{\normalsize\it
Department of Physics and Astronomy, Michigan State University }
\centerline{\normalsize\it
East Lansing, Michigan 48824 , USA }

\vspace{0.4cm}
\raggedbottom
\setcounter{page}{1}
\relax

\begin{abstract}
\noindent
We study production of $t\bar t$ via $W^{+} W^{-}$-fusion,
including the relevant backgrounds
at the proposed
Linear Collider (an $e^+ \, e^-$ collider with
$\sqrt{S}\,=\,1.5$ TeV)
in the context of a {\it Higssless}
Standard Model (SM), i.e. a nonlinear
SU(2$)_L \times$ U(1$)_Y$ chiral Lagrangian,
including dimension five
\wwtt interactions.  Deviation from the SM total cross section
can be used to constrain the coefficients of these operators
to an order of $2\times 10^{-1}$ (divided by the
cut-off scale $\Lambda = 3.1 \;$TeV) with a $95\%$ C.L..
However, there are three ways in which this sensitivity can
be improved by a factor of two.
First, by studying the deviation of the $t \bar t$ kinematics
from what is predicted by the SM; second, by polarizing the
collider electron beam; and third, by studying the polarization
of the produced top quarks.
In this way, we show that it is also possible to attempt
to dis-entangle the contributions from different anomalous
operators, isolating the form of new physics contributions.
\end{abstract}

\vspace*{3.4cm}
\end{titlepage}

\normalsize\baselineskip=15pt
\setcounter{footnote}{0}
\renewcommand{\thefootnote}{\arabic{footnote}}

\section{Introduction}
\label{intro}
\indent \indent

The Standard Model (SM) of particle physics is an amazingly
successful model, accurately predicting all available
experimental data; however the model is still incomplete.
The details of the electro-weak symmetry breaking (EWSB)
have continued to elude experimental verification.
Until there is
experimental observation of the scalar Higgs boson,
the generation of masses for the
$W$ and $Z$ bosons, and the fermions,
will remain a mystery.  If the
answer to these questions is to be found at a high energy
scale, we may still be able to gain insight into the
mechanism of the symmetry-breaking by studying the couplings
of the known vector bosons and fermions
at a some-what lower energy scale,
for deviations from
what is predicted by the SM.

In particular, the top quark
\cite{topdisc}, with its heavy mass of $\sim$ 175 GeV, the
same order as the EWSB scale
($v={(\sqrt{2}G_F)}^{-1/2}=246$\,GeV) may provide answers
to these questions.  As the heaviest of the fermions, the
top quark is unique, and may provide a useful probe of
the EWSB sector, particularly if there is a connection
between the generation of mass for the fermions and the
EWSB.  In this case, one expects some residual effects
of this mechanism could appear in accordance with
the mass hierarchy \cite{pczh,sekh,malkawi}.
Thus, new physics effects could be much more apparent
in the top quark than in the other (much lighter) fermions
of the theory.  For this reason, it is very important to study the
top quark's interactions, as they may provide information
on new physics effects \cite{kane}.

One of the reasons for the proposed Linear Collider
(LC) is to shed light upon these questions.  With a high
center of mass energy ($\sqrt{S} =$ 1.5 TeV) and a large
integrated luminosity ($L = 200 \; {\rm fb}^{-1}$),
it is expected that there will be a
few thousands of $t \bar t$ pairs and single-$t$ (or
single-$\bar t$) events produced via vector boson fusion
processes.  Thus,
the LC will allow the couplings of the longitudinally
polarized weak vector bosons
to the top quark to be very accurately determined.

In this article, we expand upon previous work \cite{top5}
by including an analysis of signal and backgrounds for
the
production of $t\bar t$ through fusion of vector bosons,
including the effects of possible anomalous dimension five
operators.  We show that at the LC
the coefficients of these operators can be measured
to order  $10^{-1}$ (divided by the cut-off scale,
$\Lambda = 4 \pi v = 3.1$ TeV).
As a comparison, the coefficients of the
next-to-leading-order (NLO) 
bosonic operators  are usually determined to about an order
of $10^{-1}$ or $1$ (divided by $\Lambda^2$)
via $V_LV_L \ra V_LV_L$
processes \cite{et,sss}.  Hence, the scattering processes
 $V_L V_L \ra t\bar t, t\ov b, \,{\rm or}\, b \bar t$ 
at high energy may provide a more sensitive probe of some 
symmetry breaking mechanisms than $V_LV_L \ra V_LV_L$,
assuming that naive dimensional analysis (NDA)
\cite{georgi} holds.

The paper is organized as follows.  In Section \ref{couplings}
we present the relevant part of the effective Lagrangian used
in our study, and provide limits on the coefficients of the
dimension five operators from partial wave perturbative
unitarity.  In Section \ref{signature}, we present a study of
the production of $t \bar t$ with missing
$p_{T_{(t \bar t)}}$ in the {\it
Higgsless} SM, including the effects of the \wwtt operators
under study.  We show how one can use the total rate of
this process, a study of the shape of rapidity distributions,
and the polarization of the top quarks to constrain the \wwtt
operators to order $10^{-1}$.  We also examine the gains one
can arrive upon by polarizing the electron or positron beams,
and find that an improvement of up to 50\% can be obtained
from a highly polarized electron beam.

\section{Dimension Five Anomalous Couplings}
\label{couplings}
\indent\indent

We wish to study new physics effects to $t\bar t$ production
at the LC in a model-independent way, using the electro-weak
chiral lagrangian (EWCL).  In the EWCL, the SU(2$)_L \times$ U(1$)_Y$
gauge symmetry is realized non-linearly, and a scalar Higgs boson
is not required for a gauge-invariant theory \cite{pczh,sekh,malkawi}.
Thus, we may study the couplings of the top quark to the gauge bosons
without assuming that a Higgs boson exists.  For this reason, we refer
to the theory without any ``new physics'' effects included as the
{\it Higgsless} SM.

We include terms of dimension five in our
effective lagrangian, and thus
NDA \cite{georgi} requires a cut-off scale,
$\Lambda$, below which the effective theory is valid.  This
scale could be identified with the lowest new heavy mass scale,
or something around $4 \pi v \simeq 3.1$ TeV if no new resonances
exist below $\Lambda$.  For the purposes of our study, we will
assume $\Lambda = 3.1$ TeV, and obtain constraints on the coefficients
of the dimension five operators.

As discussed previously \cite{top5},
there are 19 independent dimension five operators
(not including flavor-changing neutral current operators)
that
involve the top quark and the gauge bosons in the nonlinear
chiral Lagrangian, fourteen of which can contribute
to the $W_L^+ W_L^- \ra t \bar t$ process.  However, in the
expansion in powers of $E$ (the energy of the
$t\bar t$ system) of the helicity amplitudes, only seven
actually contribute to the leading terms; all the others
contribute at most to terms that are two powers below the
leading ones.  This means that in the high energy region,
where the longitudinal components of $W$ and $Z$ play the
leading role, the effects of these operators are more likely
to be observed.  In the leading terms in the
$E$ expansion, there are two types of
contribution to the partial waves of the
$W_L^+ W_L^- \ra t \bar t$
amplitudes: the S-wave and the P-wave.  Except for the
{\it derivative-on-fermion} operator \cite{top5},
every operator contributes
to either the S-wave or P-wave, but not to both.

Based on the previous discussion, and without any loss of
generality, we will simplify our study of anomalous effects of
dimension five operators on the production of $t \bar t$ pairs
at the LC by considering the two \wwtt contact operators
of our previous work \cite{top5}.  These two operators can
serve as good representatives of each class of operators,
the scalar \wwtt coupling
(i.e. $O_{g {\cal W}{\cal W}}^{(5)}$)
for those that contribute to the
S-wave, and the tensor 
\wwtt coupling
(i.e. $O_{\sigma {\cal W}{\cal W}}^{(5)})$)
for those that contribute to
the P-wave.

Therefore, the part of the
effective non-linear SU(2$)_L \times$ U(1$)_Y$ chiral
Lagrangian that is relevant for our study is:

\begin{eqnarray}
{\cal L}_{eff}&=& {\cal L}^{(4)} + 
O_{g {\cal W}{\cal W}}^{(5)}
+ O_{\sigma {\cal W}{\cal W}}^{(5)} \nonumber \\
&&= \, i\overline{t}\gamma^{\mu}\left ( \partial_{\mu}
 +i\frac{2s_w^2}{3}{\cal A}_{\mu}\right) t
+i\ov {b}\gamma^{\mu}\left (\partial_{\mu}-i\frac{s_w^2}{3}
{\cal A}_{\mu}\right ) b\nonumber \\
&&- \left (\frac{1}{2}-\frac{2s_w^2}{3}\right)
\ov {t_{L}} \gamma^{\mu} t_{L}{{\cal Z}_{\mu}}
 + \frac{2s_w^2}{3} \ov {{t}_{R}}
\gamma^{\mu} t_{R}{{\cal Z}_{\mu}} \nonumber \\
&&+\left( \frac{1}{2}-\frac{s_w^2}{3}\right)
\ov {b_{L}}\gamma^{\mu} b_{L}{{\cal Z}_{\mu}}
-\frac{s_w^2}{3}\overline{b_{R}}\gamma^{\mu} b_{R}
{{\cal Z}_{\mu}}\nonumber \\
&&-\frac{1}{\sqrt{2}}\ov {{t}_{L}} \gamma^{\mu} b_{L}
{{\cal W}_{\mu}^+}-\frac{1}{\sqrt{2}}
\ov {{b}_{L}}\gamma^{\mu}t_{L}{{\cal W}_{\mu}^-} -
m_t \overline{t} t -m_b \overline{b} b \nonumber \\ 
&& + {\frac {a_1}{\Lambda}}
\bar t  t {\cal W}_{\mu}^{+} {\cal W}^{- \mu } +
{\frac {a_2}{\Lambda}}
i \bar t  {\sigma}^{\mu\nu} t {\cal W}_{\mu}^{+} 
{\cal W}^{-}_{\nu }\; ,
\label{lagra} 
\end{eqnarray}

\noindent
where $a_1$ is the parameter which characterizes the scalar
\wwtt coupling, and $a_2$ characterizes the tensor \wwtt
coupling.

The leading terms in the $E$ expansion of the
$W^+ W^- \ra t \bar t$ helicity amplitudes
(including both the contributions from the {\it Higgsless}
SM and the anomalous dimension five terms) are:
\begin{eqnarray}
T_{++} =&& {{{m_t}\,E}\over {{v^2}}} -
{\, {{2 E^3}\over {v^2 \Lambda}} \left( {\it a_{1}}
+ {a_{2}}\,{\cos {\theta}} \right) } \nonumber \\
T_{--} =&& -T_{++} \; ,  \nonumber \\
T_{+-} =&& m_t^2 \cot{\frac{\theta}{2}} +
{{4\,E^2\,m_t\,\sin {\theta}}\over
{v^2 \Lambda}}\,a_2 \label{amplitudes} \\
T_{-+} =&& m_b^2 \cot{\frac{\theta}{2}} +
\frac{4\,E^2\,m_t\,\sin {\theta}}{v^2 \Lambda}
\,a_2   \nonumber   .
\end{eqnarray}

As mentioned before, NDA gives an estimate for the
scale $\Lambda$ to be of order $4 \pi v$ and the
coefficients $a_{1,2}$ to be of order
one\footnote{
NDA counts $\Sigma$ as $\Lambda^0$, 
$D_\mu$ as $\frac{1}{\Lambda}$, and fermion fields as 
$\frac{1}{v\sqrt{\Lambda}}$. Hence, ${\cal W}^{\pm}$, 
${\cal Z}$ and ${\cal A}$ are also counted as 
$\frac{1}{\Lambda}$. After this counting,
one should multiply the result by 
$v^2 \Lambda^2$. Notice that up to the order of intent,
the kinetic term of the gauge boson fields and the mass
term of the fermion fields are two 
exceptions to the NDA, and are of order $\Lambda^0$.
}.
Given the
strong dependance on the energy of the process
($E^3$) it is natural to ask for the limits on
the strength of the couplings to prevent violation
of perturbative unitarity.

\subsection{Constraints from Partial Wave Unitarity}
\indent\indent


As can be seen from Eq. (\ref{amplitudes}), the coupling
$a_1$ contributes
only to the S-wave, whereas $a_2$ contributes to the P-wave. 
Applying the method given in Refs. \cite{jacob,CFH} for 
partial wave analysis, we consider the leading contributions 
(that grow with any power of $E$) to the coupled channel
matrices for $J=0$ and $J=1$ in order to obtain perturbative
unitarity constraints on $a_1$ and $a_2$, respectively.
The best
constraints will come from the highest eigenvalues. 
Writing the coupled channels in the order \cite{CFH} 
$t(+) {\bar t}(+)$, $t(-) {\bar t}(-)$ and $W^+ W^-$, 
the $J=0$ coupled channel partial wave matrix is
\begin{equation}
{\cal M}_0 = \;  A \;  \left[ 
\begin{array}{ccc}
 0  &  0  & -1 \\
 0  &  0  &  1 \\
-1  &  1  &  0
\end{array} \right] \; ,  \nonumber
\end{equation} 
where,
\begin{equation}
A = \frac{a_1 E^3}{8 \pi v^2 \Lambda} -
\frac{m_t E}{16 \pi v^2} \;, \nonumber
\end{equation} 
and the largest eigenvalues are $\pm \sqrt{2} A$. 
Similarly, if we write the coupled channels in the
order \cite{CFH} $t(+) {\bar t}(-)$,
$t(+) {\bar t}(+)$, $t(-) {\bar t}(-)$, $W^+ W^-$, 
and $t(-) {\bar t}(+)$,
then the coupled channel partial wave matrix for $J=1$ is
\begin{equation}
{\cal M}_1 = \;  \left[ \begin{array}{ccccc}
0 & 0  &  0  &  C  &  0 \\
0 & 0  &  0  &  B  &  0 \\
0 & 0  &  0  & -B  &  0 \\
C & B  & -B  &  0  & -C \\
0 & 0  &  0  & -C  &  0
\end{array} \right] 
\; ,  \nonumber
\end{equation}
where,
\begin{eqnarray} B &=& -\frac{a_2
E^3}{24 \pi v^2 \Lambda} \nonumber\\
C &=& \frac{a_2 m_t E^2}{\sqrt{2}\, 6
\pi v^2 \Lambda} , \nonumber 
\end{eqnarray}
and the largest
eigenvalues are $\pm \sqrt{2 B^2 + 2 C^2}$. 


{}From the largest eigenvalues, we derive the most stringent
constraints by requiring the magnitude of each eigenvalue
to be less than unity.  The resulting bounds are
\begin{eqnarray} |a_1| &{\leq}& \sqrt{2}\,
4 \pi \frac{v^2 \Lambda}{E^3}
\left( 1+\frac{\sqrt{2} m_t E}{16 \pi v^2} \right)
\simeq 1 \, , \nonumber \\
|a_2| &{\leq}& \frac{24 \pi v^2 \Lambda}{E^2
\sqrt{2 E^2+16 m_t^2} } \simeq 2.8 \; ,
\end{eqnarray}
where we have used $\Lambda=3.1$\,TeV and $E=1.5$\,TeV for a 175\,GeV 
top quark.
Hence, in our numerical analysis, we restrict the magnitude
of the anomalous couplings $a_{1,2}$ to be within 1.0
so that the tree level perturbative calculation
does not violate the unitarity condition.

\newpage

\section{ Signature of the Anomalous Couplings:
$t \bar t$ with missing $p_{T_{(t \bar t)}}$}
\label{signature}
\indent \indent

\begin{figure}
\centerline{\hbox{
\psfig{figure=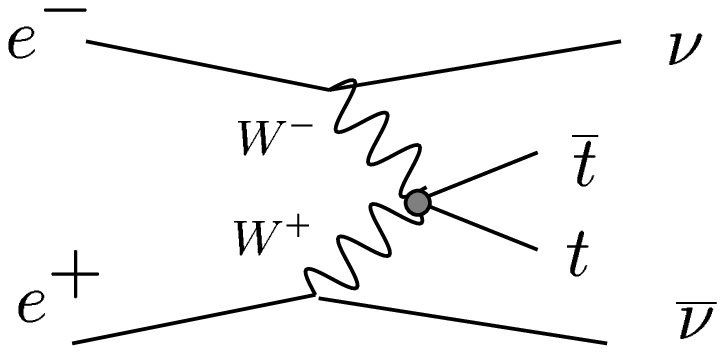,height=1.0in}}}
\caption{ Diagrams for $e^- e^+ \ra \nu {\ov \nu} t \bar t$ through
anomalous \wwtt couplings.} 
\label{diagan}
\end{figure}

The \wwtt coupling, appearing in the $W^+ W^-$ fusion
process will generate a $t \bar t$ pair in the central region
of the detector with missing
transverse momentum carried away by the neutrinos produced
by the charged leptonic current [cf. Fig. \ref{diagan}].

As this process also appears in the
{\it Higgsless} SM, what we hope to observe
is actually an interference between the diagrams containing
the anomalous couplings
and those from {\it Higgsless} SM.  Let us analyze the latter for
the moment, and then examine the effects of the \wwtt operators
in Section \ref{wwtteffect}.

\subsection{ $t \bar t$ with missing $p_{T_{(t \bar t)}}$ in
the {\it Higgsless} SM}
\indent \indent

The {\it Higgsless SM} prediction
for the process $e^+ e^- \ra t {\bar t}\, $
with
missing $p_{T_{(t \bar t)}}$
consists dominantly of two subprocesses: one
in which the
missing $p_{T_{(t \bar t)}}$ is carried by two neutrinos, and
the other where it is carried by a particle (i.e., a hard photon)
which escapes detection.

\begin{figure}
\centerline{\hbox{
\psfig{figure=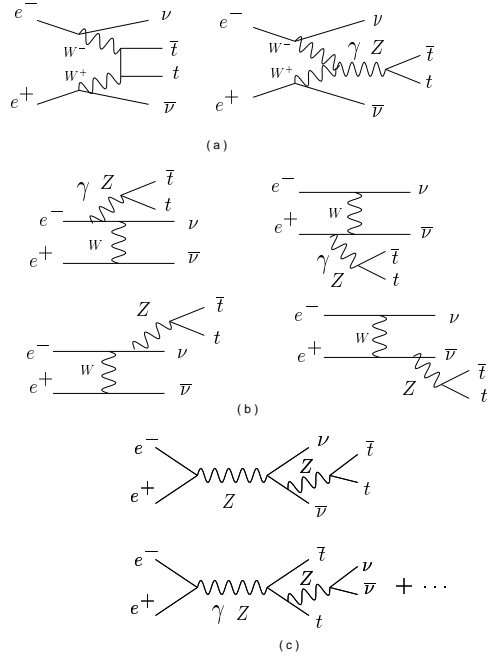,height=3.5in}}}
\caption{ Representative diagrams for {\it Higgsless} SM
$e^- e^+ \ra \nu {\ov \nu} t \bar t$:
(a) $W^+ W^-$ fusion, (b) $W$ exchange, and (c) $e^- e^+$ anihilation.} 
\label{diagsm}
\end{figure}

In the {\it Higgsless} SM there are 19 tree level diagrams for
the process $e^- e^+ \ra \nu {\ov \nu} t \bar t$.
There are three
$W^+ W^-$ {\it fusion} diagrams which,
at high energies\footnote{
$\sqrt{S}=1.5$ TeV $\gg M_Z,\, M_W$, the masses of the
$Z$ and $W$ bosons.}
are the most important
contributions, along with six $W$ {\it exchange} diagrams of a
lower magnitude [cf. Fig.  \ref{diagsm}(a) and (b) respectively].
(These nine diagrams form a gauge invariant subset).
The set of 10 diagrams in Fig. 2(c) contribute to less than about
one percent of the total $t {\bar t} \nu {\bar \nu}$ rate 
after imposing the kinematic cuts (cf. Eq. (\ref{cutseq}))
for suppressing the 
$\gamma t \bar t$ background\footnote{In principle, one can impose a
cut on the invariant mass $M_{\rm inv}(\nu \bar \nu)$
of the invisible particle system (i.e. the $\nu \bar \nu$ pair) to further
suppress the background at the expense of some signal rate.
However, in our analysis, we shall not impose such a cut to retain
enough statistics.}.

\begin{figure}
\centerline{\hbox{
\psfig{figure=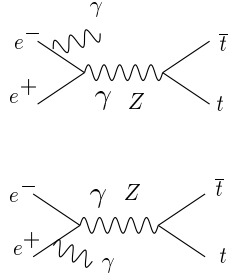,height=1.5in}}}
\caption{ Diagrams for the ISR subprocess,
$e^- e^+ \ra \gamma t \bar t \;$.} 
\label{diagpho}
\end{figure}

Since our signal consists of a $t \bar t$
pair with missing transverse momentum (carried by the neutrinos),
we also must consider the background process (at order $\alpha_{em}^2$)
$e^- e^+ \ra \gamma t \bar t$ [cf. Fig.  \ref{diagpho}],
where the photon escapes the range
covered by the detector\footnote{One of the
current design proposals for the
LC estimates that 0.15 rad about the beam
axis will not be covered
by the detector \cite{LC}.}, but is
sufficiently hard to generate the required missing $p_{T_{(t \bar t)}}$.
The pictured diagrams correspond to initial state radiation (ISR).
There are also contributions due to final state radiation (FSR),
where a photon is radiated from the $t$ or $\bar t$ in the final
state, but these contributions are suppressed by the heavy
top mass and thus provide a negligible effect.

Using a Monte-Carlo program to calculate the total
cross section for the $e^- e^+ \ra \nu {\ov \nu} t \bar t$
process, we compare the results with
the ones obtained in Ref. \cite{top5}, in which the effective
$W$ approximation method was applied and a total cross section
{$\sigma_{{W^+}_L {W^-}_L \ra t \bar t} = 2$~fb} was obtained.
In that calculation only the longitudinal components of the
massive $W$ bosons were considered, as well as the leading terms
in powers of $E$ for the helicity amplitudes
${W^+}_L {W^-}_L \ra t \bar t$.
By applying the same requirements as used
in the effective $W$ study,
we can compare our exact calculations to see
how well the effective $W$
approximation works in this process.  
Namely, we require the invariant mass of the $t \bar t$ system
to be $M_{t \bar t} > 500$~GeV and
the rapidity and the transverse momentum of $t$ and $\bar t$ 
to be $|y|\leq 2$ and $p_T \geq 20$\,GeV, respectively.
{}From the complete matrix element calculation,
we obtain a similar value of 2.2 fb for the total cross section.
Comparing this with the $2$~fb obtained from the effective
$W$ approximation,
we conclude that the effective $W$ approximation provides an
accurate estimate of the total cross section
in this kinematic region to about 10\%.

In order to suppress the background from the ISR
subprocess of $e^+ e^- \ra \gamma t \bar t$,
we have found
it essential to require that the $t\bar t$ system have
a minimum of transverse momentum, $p_{T_{(t \bar t)}} \geq 20$\,GeV.
Because this ISR subprocess dominantly produces
photons approximately co-linear with the beam axis, this constraint
will remove much of the background.  Any photon with
polar angle $\theta$ less than 0.15 rad and energy more
than $20/0.15 \simeq 130$ GeV will not be removed by this
cut.  Roughly speaking, only $10\%$ of the initial state
radiation carries this (or more) energy \cite{ISR}.
Thus we are effectively cutting out most of the ISR background,
which nonetheless remains the order of the
$W^+ W^-$ fusion {\it Higgsless} SM cross
section ($\sigma_{\gamma t \bar t} = 1.8$~fb).

After imposing the minimal set of constraints:
\begin{eqnarray}
|y_t|,\,|y_{\bar t}|\, & \leq  & 2 \, ,\nonumber \\
p_{T_{(t)}}, p_{T_{(\bar t)}} & \geq & 20 \,{\rm GeV} \, , \nonumber \\
p_{T_{(t \bar t)}} &\geq & 20 \,{\rm GeV}, \label{cutseq}
\end{eqnarray}
we find that the cross section for
$e^+ e^- \ra \nu {\bar \nu}t \bar t$ is about 4.0 fb in the
{\it Higgsless} SM.
This indicates that the effective $W$ approximation
used in \cite{top5} estimated the total rate for $t \bar t$
production via $W^+ W^-$ fusion at the LC by about a factor
of 2 too small\footnote{Roughly speaking, only half of
the ${t \bar t}$ pairs meet the requirement
$M_{t \bar t} > 500$GeV associated with the effective W
approximation.
The transverse $W$ boson contributions were not included
in the previous work \cite{top5}, but are
included here using the exact scattering amplitudes.} .
However, as we shall show, this will not have a large effect
on the limits which can be obtained on $a_{1,2}$ by studying
the total rate for $t \bar t$ with missing $p_{T_{(t \bar t)}}$
at the LC, particularly when the ISR contribution to the
background is included.

\begin{figure}
\centerline{\hbox{
\psfig{figure=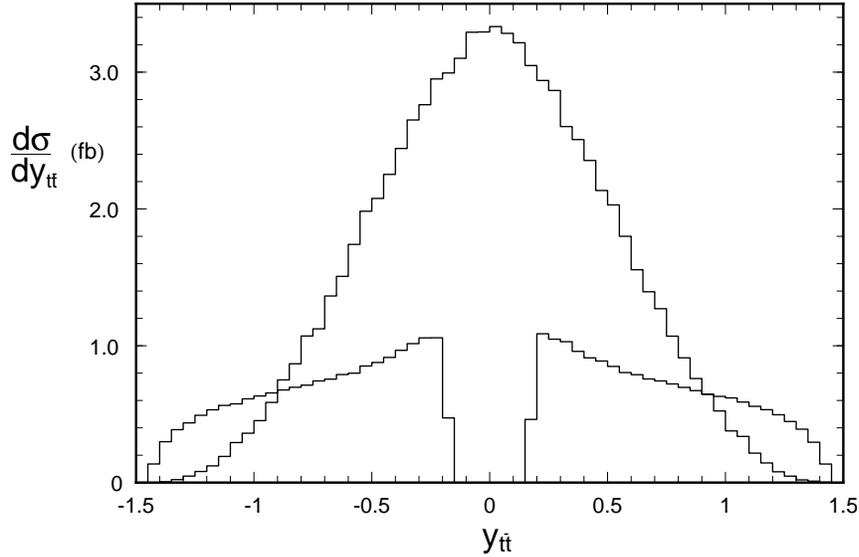,height=3.0in}}}
\caption{ Distribution of $y_{t \bar t}$
for {\it Higgsless} SM
$e^- e^+ \ra \nu {\ov \nu} t \bar t$ (upper curve), and for
$e^- e^+ \ra \gamma t \bar t$ (lower curve).} 
\label{yttsm}
\end{figure}

The two type of processes, $\nu {\bar \nu}t \bar t$ by $W^+ W^-$
fusion, and $\gamma t \bar t$ by $e^+ e^-$ anihilation, have
different distributions in several kinematical variables.
For instance,
in Fig. \ref{yttsm} we show
the
rapidity ($y_{t \bar t}$) of
the $t \bar t$ system for each subprocess.
For the ISR process, which is a two-to-three process, the
rapidity has a minimum absolute value of
about 0.18; this means that the $t \bar t$ pair must have a
minimum thrust toward the beam axis.
This can be easily understood from the fact that
the emitted photon, which must be directed close to
the beam axis in order to escape detection, 
must carry enough energy in order to generate the
required missing $p_{T_{(t \bar t)}}$,
and consequently the $t \bar t$ system is boosted.
On the other hand, the $W^+ W^-$ fusion mechanism prefers
the $t$ and $\bar t$ to be produced
in the rapidity region close to zero.

\begin{figure}
\centerline{\hbox{
\psfig{figure=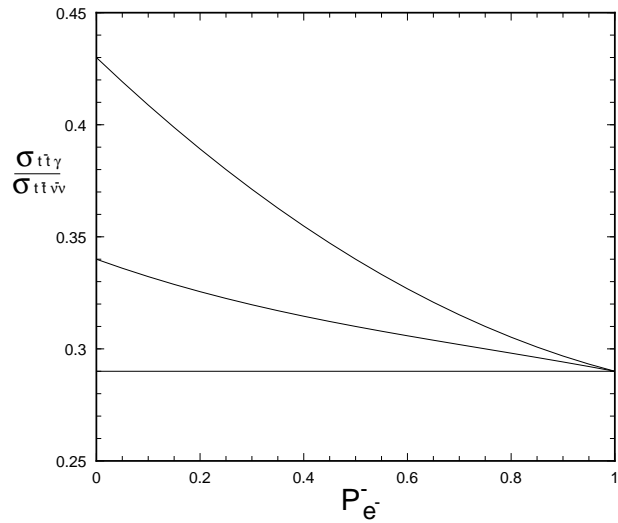,height=2.8 in}}}
\vspace{1.0in}
\caption{ Ratio of $\sigma_{e^- e^+ \ra t \bar t \gamma}$
and $\sigma_{e^- e^+ \ra \nu {\ov \nu} t \bar t}$ as a
function of the left handed polarization of the electron
beam and for three
different values of $P^+_{e^+}$ right handed polarization
of the positron beam: 0, 0.5 and 1 for the upper, middle
and lower curves respectively.} 
\label{pole}
\end{figure}

Another important difference between these processes is
their dependence on the polarization of the initial
electron and positron beams.  For $W^+ W^-$ fusion the
only non-zero contribution comes from a purely
left-handed electron and a purely right-handed
positron\footnote{This is due to the left-handed nature
of the $W$ coupling and
the fact that the electron can be treated as massless at
$\sqrt{S} = 1.5$ TeV.}.
The rate for polarized $e^+$ or $e^-$ beams can
be written in terms of the rate for unpolarized beams by
using the relation,
\begin{eqnarray}
\sigma_{e^- e^+ \ra \nu {\ov \nu} t \bar t} \,=\,
\sigma^{(0)}_{e^- e^+ \ra \nu {\ov \nu} t \bar t}
\times (1+P^-_{e^-}) \times (1+P^+_{e^+})
\end{eqnarray}
\noindent
where $\sigma^{(0)}_{e^- e^+ \ra \nu {\ov \nu} t \bar t}$ is
the unpolarized beam rate (determined to be 4 fb after imposing
the minimal cuts described above), and $P^-_{e^-}$ ($P^+_{e^+}$)
are the fractional left (right) polarization of the electron
(positron) beams respectively.  On the other hand,
for $t \bar t \gamma$ production the vector
coupling also allows for contribution from a right (left)
handed electron (positron).  Therefore, to
enhance the signal-to-background ratio we should consider
the possibility of enhancing the $W^+ W^-$ fusion with an
$e^-$ ($e^+$) beam with some degree of left (right)
handed polarization.  In Fig. \ref{pole} we show
the ratio of the total cross sections of these two
processes as a function of the left-handed polarization
of the electron beam, and for three different values of
right-handed polarization of the positron beam.

\begin{figure}
\centerline{\hbox{
\psfig{figure=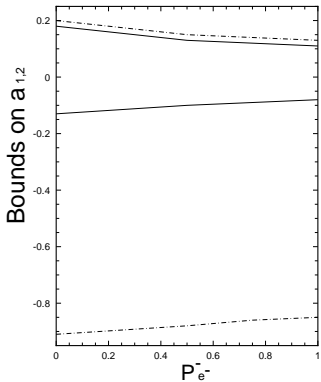,height=2.0in}}}
\caption{ Improvement on the upper and lower bounds of
$a_1$ and $a_2$ (solid and dot-dashed lines
respectively) for an unpolarized positron
beam but a left-handed polarized electron
beam.} 
\label{boundspol}
\end{figure}

Apart from the expected reduction of
$\sigma_{e^- e^+ \ra t \bar t \gamma}$ relative to
the size of $\sigma_{e^- e^+ \ra \nu {\ov \nu} t \bar t}$
for higher degrees of polarization, there are two things
that can be noticed from Fig. \ref{pole}. First,
for higher degrees of (left-handed)
polarization of the electron
beam there is no substantial reduction in the ratio
${\sigma_{e^- e^+ \ra t \bar t \gamma}}/
{\sigma_{e^- e^+ \ra \nu {\ov \nu} t \bar t}}$
if we also increase the positron's right-handed
polarization. Second, for the case of an unpolarized
positron beam the rate of reduction of the ratio falls off
as one polarizes the electron beam more and more.
In other words,
there is more progress in the relative reduction of
${\sigma_{e^- e^+ \ra t \bar t \gamma}}$ as one
sets the electron's polarization from 0 to 0.5 than
from 0.5 to 1.0.  As a consequence of this, the
best improvements in the bounds of the anomalous
couplings $a_1$ and $a_2$ take place in the lower
degrees of polarization (assuming that
no significant deviation from the SM prediction is observed).
In Fig. \ref{boundspol} we show the improvement of
the bounds for $a_1$ and $a_2$ depending on the
polarization of the electron beam (assuming an unpolarized
positron beam).  From this study we see that an
improvement of about 43\% for $a_1$ and 11\% for $a_2$
results when one considers the
constraint obtained from an unpolarized beam compared to
that which is possible from a completely polarized beam.
These results lead us to conclude that a polarized electron
beam can be an important and
useful tool in probing this type of new physics
effect at the LC.

\begin{figure}
\centerline{\hbox{
\psfig{figure=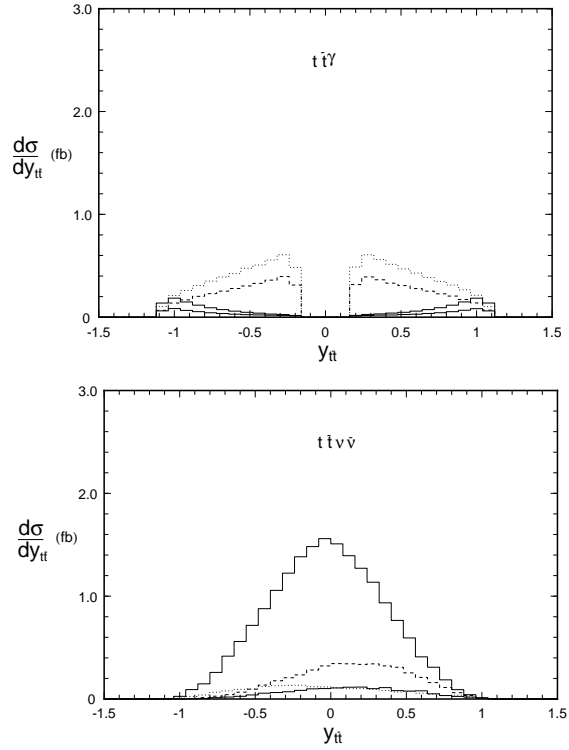,height=4.0in}}}
\caption{ The distribution of $y_{t \bar t}$ for
{\it Higgsless} SM
$e^- e^+ \ra \nu {\ov \nu} t \bar t$ (lower figure), and for
$e^- e^+ \ra \gamma t \bar t$ (upper figure), for various 
$t$ and $\bar t$ helicities.  The
solid lines indicate parallel
helicities, $t(+) {\bar t}(+)$ (upper curves)
and $t(-) {\bar t}(-)$ (lower curves).
The
dashed line shows antiparallel helicities,
$t(+) {\bar t}(-)$.  The dotted line
represents helicity combination $t(-) {\bar t}(+)$.} 
\label{yttsmpol}
\end{figure}

One more feature that distinguishes the two subprocesses
is the polarization of the $t$ and $\bar t$ quarks.
Should it turn out to be experimentally feasible to measure
the polarizations of the $t$ and $\bar t$, this information
can also be used to separate the two subprocesses.
In $W^+ W^-$ fusion we find the parallel helicities
(equal sign) final states to dominate
over the antiparallel ones.  On the other hand,
for the ISR sub-process we find that the opposite is true.  
 To illustrate this point, we show in 
 Fig. \ref{yttsmpol} the contributions to the rapidity
 distributions from the various $t$ and $\bar t$ polarization
 combinations.

\subsection{ Effects from the anomalous \wwtt couplings }
\label{wwtteffect}
\indent \indent

\begin{figure}
\centerline{\hbox{
\psfig{figure=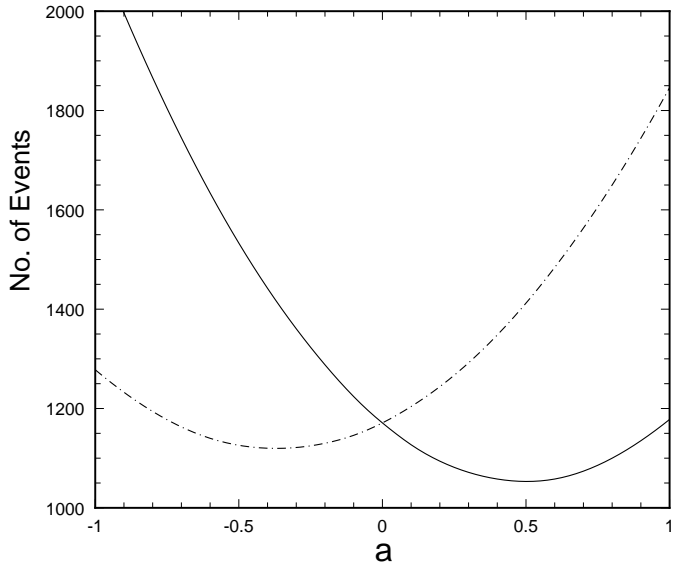,height=3.0in}}}
\caption{ Number of $W^+ W^-$ fusion $t \bar t$ pairs as a function
of the anomalous couplings $a_{1}$ (solid line) and
$a_{2}$ (dash-dot line).  The point $a_1 = a_2 = 0$
corresponds to the prediction of the {\it Higgsless} SM.}
\label{events}
\end{figure}

As demonstrated in \cite{top5}, the anomalous couplings may
be discovered or constrained by examining their effect upon the
total production rate of $t \bar t$ pairs at the LC.
In Fig. \ref{events} we show the total number of $W^+ W^-$ fusion
events expected
at the LC with integrated luminosity $L = 200 \; {\rm fb}^{-1}$,
as a function
of the anomalous operator couplings, $a_1$ and $a_2$.
We can use these
results to obtain bounds on $a_1$ and
$a_2$ at the 95\%~C.L., provided
no new physics effects are observed at the LC.  In Table \ref{bounds},
we present the constraints on $a_1$ and $a_2$
from the total rate estimated in \cite{top5},
as well as the more realistic bounds obtained in this work, including
the ISR process $t \bar t \gamma$ discussed above,
and the intrinsic {\it Higssless} SM background processes.
We find that the
two results are
in agreement to within a factor of two,
though
the more realistic estimation
does suffer in some cases when the $t \bar t \gamma$ background is
included.

\begin{figure}
\centerline{\hbox{
\psfig{figure=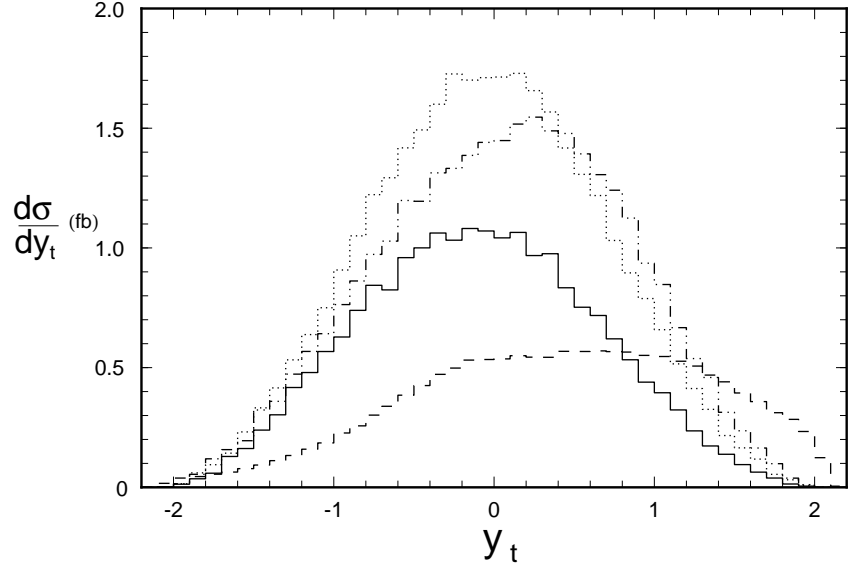,height=3.0in}}}
\caption{ Distribution of rapidity of $t$ for the
{\it Higgsless} SM
$W^+ W^-$ fusion process, the
$t \bar t \gamma$ process, and the $W^+ W^-$ fusion process including
one value of $a_1$ and $a_2$.  The solid line corresponds to
{\it Higgsless} SM $e^- e^+ \ra \nu {\ov \nu} t \bar t$.  The dashed
line is $e^- e^+ \ra \gamma t \bar t$.
The dashed-dotted line is $e^- e^+ \ra \nu {\ov \nu} t \bar t$
with $a_{2} = 0.6$ and the dotted line
corresponds to $e^- e^+ \ra \nu {\ov \nu} t \bar t$ with
$a_{1} = -0.4$.}
\label{ytop}
\end{figure}

\begin{figure}
\centerline{\hbox{
\psfig{figure=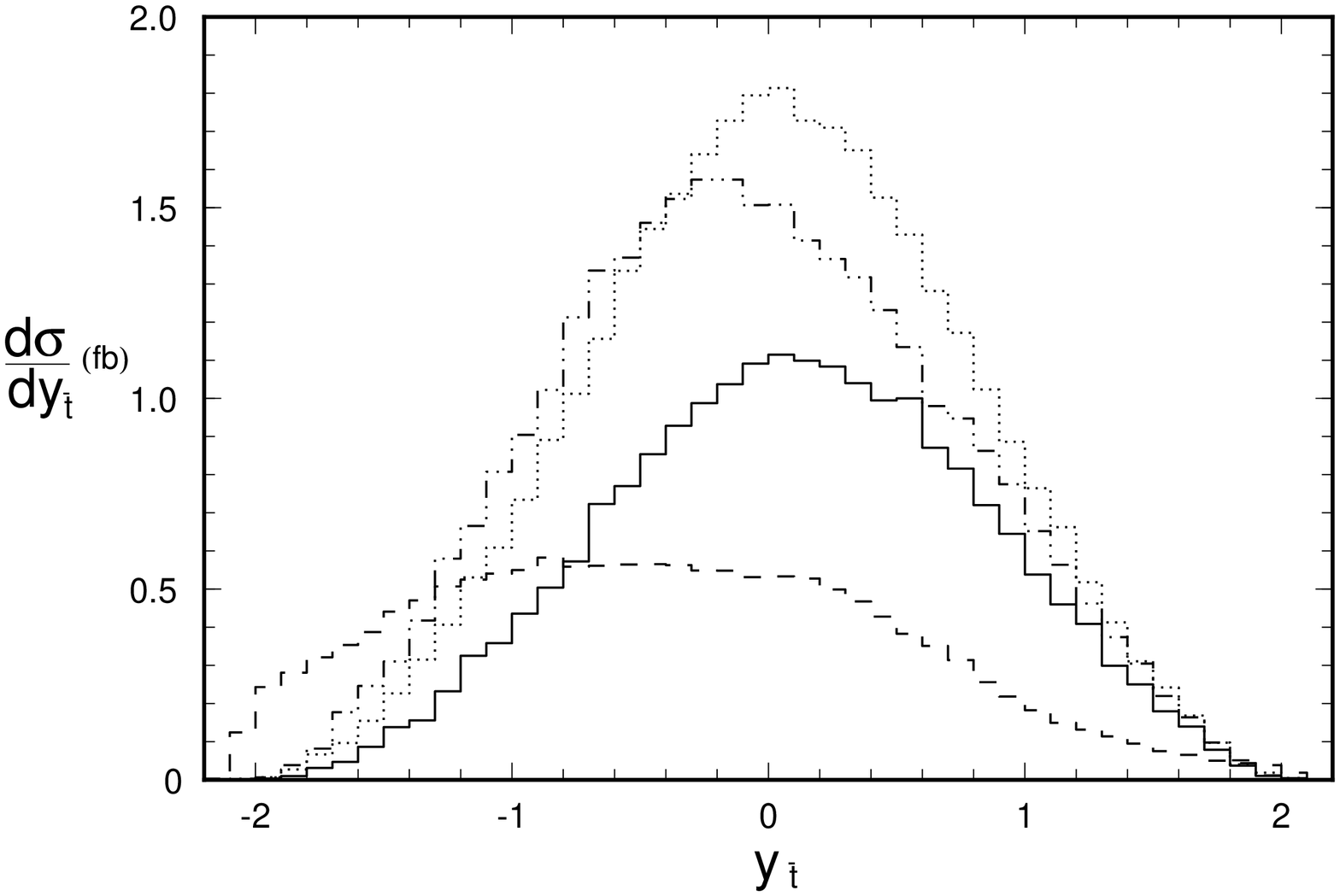,height=3.0in}}}
\caption{ Same as Fig. 9, but for $y_{\bar t}$.}
\label{ytbar}
\end{figure}

\begin{figure}
\centerline{\hbox{
\psfig{figure=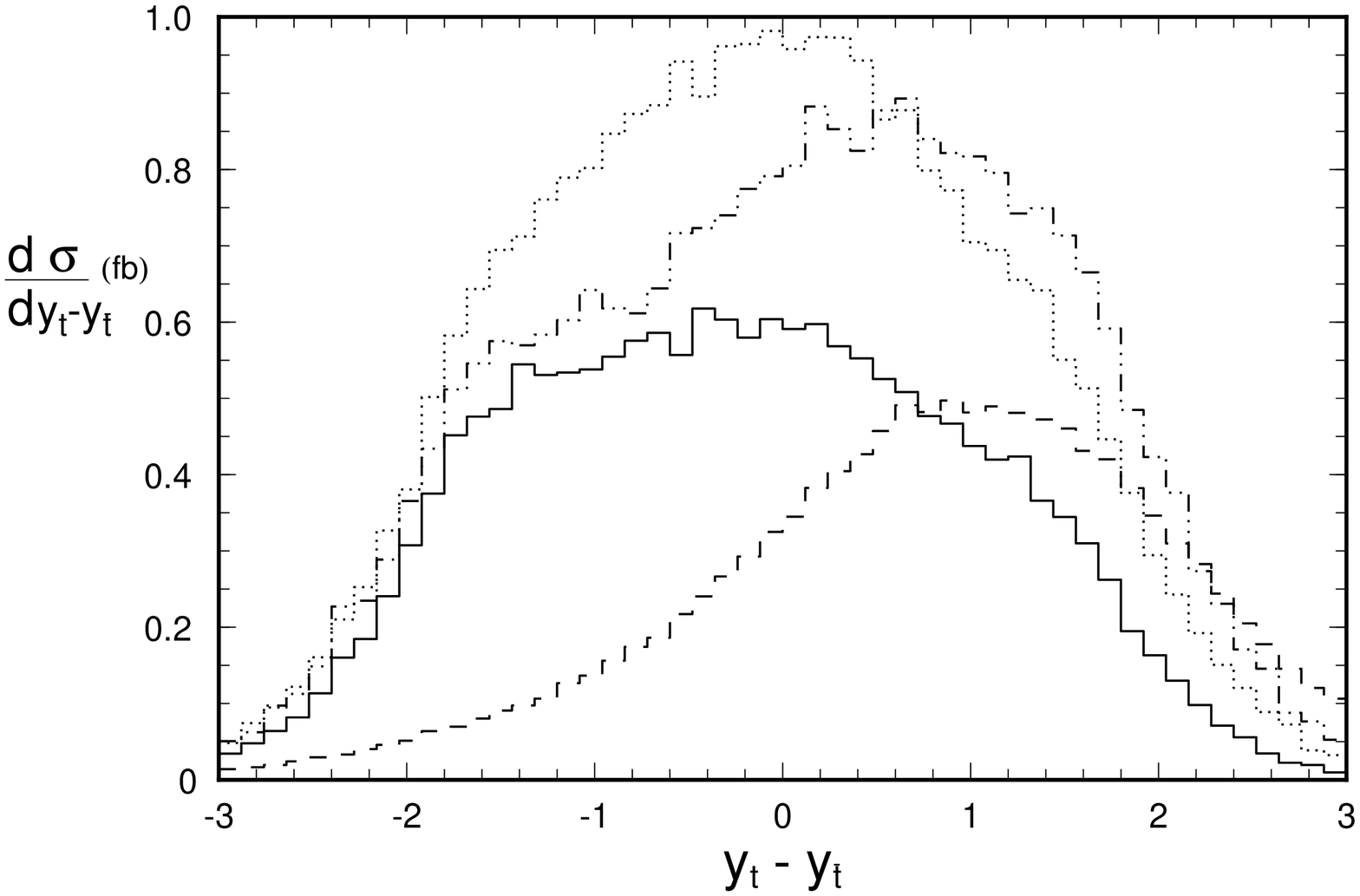,height=3.0in}}}
\caption{ Same as Fig. 9, but for $y_t - y_{\bar t}$.}
\label{rapdiff}
\end{figure}

However, as we have discussed above, the background from $t \bar t \gamma$
has different rapidity distributions from the
$W^+ W^-$ fusion process
we wish to study.
In Figs. \ref{ytop}, \ref{ytbar}, and \ref{rapdiff}
we show that the effects of \wwtt operators
can modify the shape of
these distributions.  Thus, by examining the effect on the
shape of the rapidity distributions,
we find that it is possible to
improve the bounds obtained by studying
the total cross section.  Because
the contribution of the scalar operator ($a_1$)
is largely S-wave, as
is the SM contribution, the scalar
operator does not have a large effect
on the rapidity distributions of $t$, $\bar t$,
or the $t \bar t$ system.
Thus we find that the bounds on $a_1$
improve only slightly.  For the
tensor operator (which contributes to the P-wave)
we find a larger effect, and there is a significant
improvement on the lower constraint
on $a_2$.  In particular, we
find that the effect of the tensor
operator is to shift the rapidity
distribution of the $t$ and $\bar t$
in opposite directions [cf.
Figs. \ref{ytop}, \ref{ytbar},
and \ref{rapdiff}].  For
this reason it is useful to look
at the distribution of
$y_t - y_{\bar t}$ in order to constrain $a_2$.

\begin{table}[htbp]
\begin{center}
\vskip -0.06in
\begin{tabular}{|l||c||c|}\hline \hline
Quantity &   bounds for $a_1$ & bounds for $a_2$  \\ \hline
Events($W^+ W^- \ra t\bar t$) & $-0.06 \leq a_1 \leq 0.07$ & 
$-0.56 \leq a_2 \leq 0.24$ \\ \hline
Events($e^+ e^- \ra t{\bar t} \nu \bar \nu$) &
$-0.13 \leq a_1 \leq 0.18$ & 
$-0.9 \leq a_2 \leq 0.2$ \\ \hline 
$\chi^2(y_{t\bar t})$  & $-0.10 \leq a_1 \leq 0.12$ & 
$-0.78 \leq a_2 \leq 0.18$  \\ \hline
$\chi^2(y_{t}-y_{\bar t})$ & $-0.18 \leq a_1 \leq 0.3$ & 
$-0.3 \leq a_2 \leq 0.2$ \\ \hline \hline
\end{tabular}
\end{center}
\vskip 0.08in
\caption{The $95\%$ C.L. limits on $a_1$ and $a_2$ obtained 
from studying the total $t \bar t$ production rate at the LC
and from $\chi^2$ analysis of the
effects on the distributions of
$y_{t \bar t}$ and $y_t - y_{\bar t}$.}
\label{bounds}
\end{table}

\begin{figure}
\centerline{\hbox{
\psfig{figure=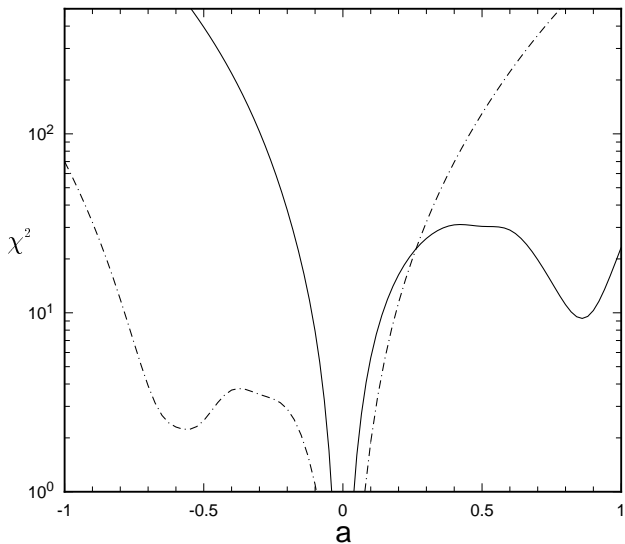,height=3.0in}}}
\caption{ 
$\chi^2$ function for $y_{t \bar t}$ as a function of $a_1$
(solid line) and $a_2$ (dash-dotted line). }
\label{yttx2}
\end{figure}

\begin{figure}
\centerline{\hbox{
\psfig{figure=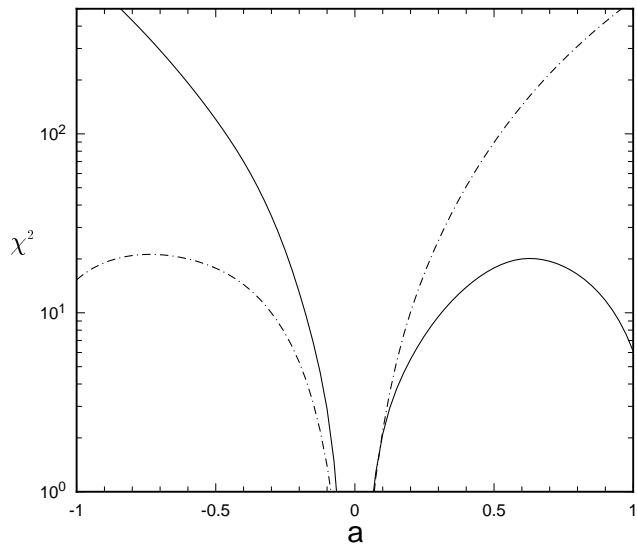,height=3.0in}}}
\caption{ 
$\chi^2$ function for $y_{t} - y_{\bar t}$.
The solid line shows the $\chi^2$ as a function of $a_1$,
while the dash-dotted line shows $\chi^2$
as a function of coupling $a_{2}$.} 
\label{ydiffx2}
\end{figure}

We analyze the $y_t - y_{\bar t}$ and $y_{t \bar t}$
distributions by
computing the $\chi^2$ deviation between the
distribution including
the anomalous effects
and that predicted by the
{\it Higssless} SM, according to the formula,
\begin{eqnarray}
\chi^2 \,=\, \Sigma^K_{j = 1}
\frac{(N^{A}_j - N^{SM}_j)^2}{N^{SM}_j} ,
\end{eqnarray}
where $K$ is the total number of bins in the histogram, $N^{A}_j$
is the number of events in bin $j$
including the anomalous effects, and $N^{SM}_j$ is the
number of events predicted by the
{\it Higgsless} SM in bin $j$. 
We find that given the statistics available at the LC,
for the $y_{t \bar t}$ analysis it is optimal to use 3
(equally sized) bins in the region
$-0.6 \leq y_{t \bar t} \leq 0.6$, and 4 bins for
the $y_t - y_{\bar t} \;$ analysis (in the region
$0 \leq y_t - y_{\bar t} \leq 2$).  In Figs. 
\ref{yttx2} and \ref{ydiffx2} we show the dependance
of $\chi^2$ on the size of the anomalous couplings.  Carrying
through this analysis and extracting the 95\% C.L.
bounds\footnote{For the three bin analysis of $y_{t \bar t}$,
a 95\% C.L. corresponds to a $\chi^2$ of 7.8, while for
the four bin analysis of $y_t - y_{\bar t}$, a $\chi^2$ of
9.8 corresponds to a 95\% deviation\cite{pdb}.}
on $a_1$ and $a_2$, we find, as expected,
that the limits on $a_1$ show a small improvement, while the
limits on $a_2$ may be improved by about a factor of 2 by considering
the $y_t - y_{\bar t}$ distribution.  These results are summarized in
Table \ref{bounds}.

The $\chi^2$ deviation function for $y_{t \bar t}$ shows an interesting
behavior at positive (negative) values of
$a_1$ ($a_2$). Instead of growing continually for greater
values of the anomalous coupling, it turns over at about
a value of 0.5 (-0.3), falls to a minimum at 0.85 (-0.6), and
then rises again as the anomalous coupling is increased.
This can easily be understood from Fig. \ref{events}, which
shows the total number of events expected
as a function of $a_1$ and $a_2$.  Since the shape
of the $y_{t \bar t}$ distribution for the
{\it Higgsless} SM and
anomalous cases are very similar, the $\chi^2$ is largely
a measure of the effect of the anomalous couplings on
the total production rate.  For positive (negative) values
of $a_1$ ($a_2$) the interference between the
{\it Higgsless} SM and
anomalous amplitudes is destructive, and thus the rate
diminishes.  However, as $a_1$ ($a_2$) becomes more positive
(negative) the anomalous amplitude begins to dominate the
{\it Higgsless} SM
one and the rate returns to the
{\it Higgsless} SM value, thus causing a local
minimum in the $\chi^2$ function.  As the magnitude of the
anomalous coupling continues to rise, the production rate
rises monotonically above the
{\it Higgsless} SM value, and thus the $\chi^2$
increases.  This feature is generic to any such analysis where
the process under study receives both
SM and anomalous
contributions to the total rate (and thus can interfere at the
amplitude level) and the distribution under study does not
under-go a large change in shape due to the effect of the
anomalous operator.

It may be possible to use the information of the rapidity
distributions under consideration to dis-entangle the
effects of the two \wwtt operators under study.  In
this way one could identify whether an observed 
\wwtt new physics effect at the LC was due to the scalar
operator or the tensor operator in Eq. (\ref{lagra}).  As
we have shown in Fig. \ref{rapdiff}, 
the $y_t - y_{\bar t}$ distribution shape
can be considerably modified by the effect
of a tensor operator,
where-as the scalar operator shows a distribution much
more like that predicted by the 
{\it Higgsless} SM.  Thus if a signal
is observed at the LC, the rapidity distribution can serve
to help identify which operator is responsible.

Moreover, there is a correlation between the $\chi^2$
analysis and the total production rate. For instance,
assume that a there are a total of 1400 events (about
200 greater than the 
{\it Higgsless} SM prediction), then according to
Fig. \ref{events} this effect is due to either $a_1$
being of order -0.3 or $a_2$ being of order 0.5.
If $a_1 = -0.3$ we expect then $\chi^2_{y_{t\bar t}}$
to be of order $10^2$ and $\chi^2_{y_{t}-y_{\bar t}}$
to be of order 30.  On the other hand,
if $a_2 = 0.5$ we also expect $\chi^2_{y_{t\bar t}}$
to be of order $10^2$, but $\chi^2_{y_{t}-y_{\bar t}}$
to be of order $90$ (3 times higher than the other case).

\subsection{Polarization of $t$ and $\bar t$}
\indent

Should it prove possible to reconstruct the polarization
information of the $t$ and $\bar t$ at the LC, we expect
that this information could be useful
in both improving the bounds on $a_{1,2}$ and in
identifying the operator responsible if a new physics
signal is observed at the LC.

In Eq. (\ref{amplitudes})
we presented the dependence on the anomalous couplings of
the process $W^+ W^- \ra t \bar t$ for the four possible
polarizations of the $t$ and $\bar t$ final state.
Since
the $t(+) \bar t(-)$ and $t(-) \bar t(+)$ matrix elements
do not depend on $a_1$, it could be possible to use their
rates to probe $a_2$ independently
from $a_1$.  In the full calculation of $e^+ e^- \ra
t \bar t \nu \bar{\nu}$, we find that this is indeed the
case.
If a large deviation is observed only
for parallel $t \bar t$ polarizations
($t(+){\bar t}(+)$ or $t(-){\bar t}(-)$),
with no corresponding
effect in the anti-parallel rate, one could thus be sure
that the tensor operator, $O_{\sigma {\cal W}{\cal W}}^{(5)}$,
is not responsible.
As discussed above, the rapidity distribution
$y_{t} - y_{\bar t}$ could further be used to identify the
scalar operator, $O_{g {\cal W}{\cal W}}^{(5)}$,
as the source of the deviation.

\begin{figure}
\centerline{\hbox{
\psfig{figure=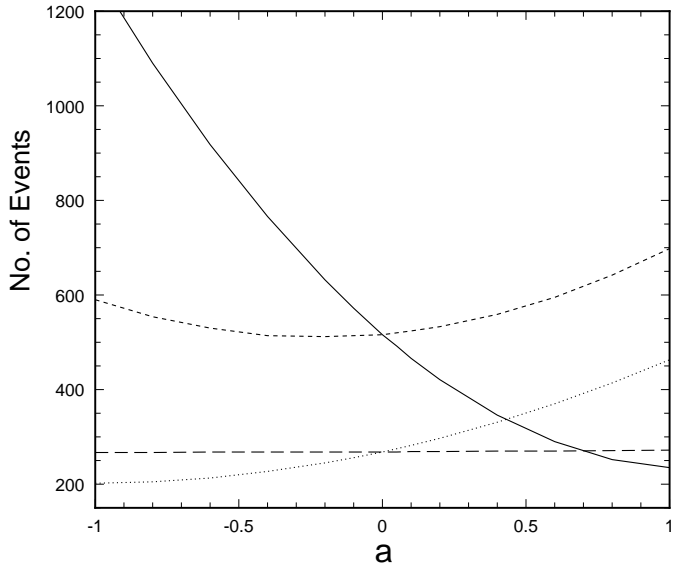,height=3.0in}}}
\caption{ Number of $t \bar t$ pairs of various
polarizations at the LC (with $L = 200 \;{\rm pb}^{-1}$
of data) as a function
of the anomalous couplings.  The
solid line is the number of $t{(+)} {\bar t}{(+)}$
produced as a function of $a_{1}$.  The
long-dashed line is the number of $t{(+)} {\bar t}{(-)}$
as a function of $a_{1}$.  The
short-dashed line is the number of $t{(+)} {\bar t}{(+)}$
as a function of the coupling $a_{2}$.  The
dotted line is the number of $t{(+)} {\bar t}{(-)}$
as a function of $a_{2}$.} 
\label{evepol}
\end{figure}

The total number of events
for various $t$ and $\bar t$ polarizations
at the LC (assuming an
integrated luminosity of $L = 200 \;{\rm fb}^{-1}$) are
shown in Fig. \ref{evepol}.  As noted above, we see that
the $t(+) \bar t(-)$ rate is independent of $a_1$.  Thus,
by studying the $t(+) \bar t(+)$ rate we find that we
can improve the bounds on $a_1$ by about a factor of 2
(here, we are considering unpolarized electron and
positron beams)
[cf. Table \ref{boundspol2}].  We also find a substantial
improvement in the lower bound on $a_2$ by studying the
$t(+) \bar t(-)$ rate, due to the 
fact that the 
{\it Higgsless} SM contribution to this channel is small,
thus making new effects from the
tensor operator more prominent.

\begin{table}[htbp]
\begin{center}
\vskip -0.06in
\begin{tabular}{|l||c||c|}\hline \hline
Quantity &   bounds for $a_1$ & bounds for $a_2$  \\ \hline
Events($e^+ e^- \ra \nu \bar \nu t(+){\bar t}(+)$) &
$-0.08 \leq a_1 \leq 0.08$ & 
$-0.85 \leq a_2 \leq 0.42$ \\ \hline
Events($e^+ e^- \ra \nu \bar \nu t(+){\bar t}(-)$) &
- & $-0.28 \leq a_2 \leq 0.20$ \\ \hline 
$\chi^2(y_{t (+)\bar t (+)})$  & $-0.08 \leq a_1 \leq 0.08$ & 
$-0.76 \leq a_2 \leq 0.4$  \\ \hline
$\chi^2(y_{t (+)\bar t (-)})$ & - & 
$-0.28 \leq a_2 \leq 0.20$ \\ \hline
$\chi^2(y_{t (+)}-y_{\bar t (+)})$ & $-0.15 \leq a_1 \leq 0.15$ & 
$-0.20 \leq a_2 \leq 0.20$ \\ \hline
$\chi^2(y_{t (+)}-y_{\bar t (-)})$ & - & 
$- \leq a_2 \leq 0.38$ \\ \hline \hline
\end{tabular}
\end{center}
\vskip 0.08in
\caption{The bounds from polarized $t$ and $\bar t$
quantities ($95\%$ CL).  Since $a_1$ is not sensitive to
the $t(+) \bar t(-)$ production, no bounds on $a_1$ are
given for that polarization.  There is no 95\% C.L.
lower bound on $a_2$ from $y_t - y_{\bar t}$.}
\label{boundspol2}
\end{table}

\begin{figure}
\centerline{\hbox{
\psfig{figure=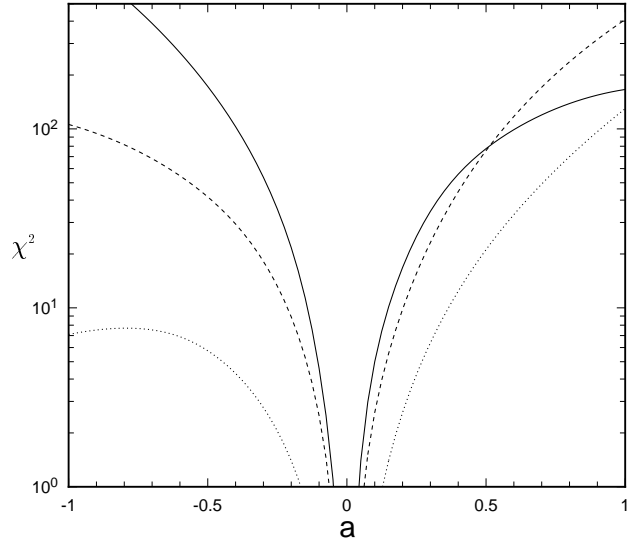,height=3.0in}}}
\caption{ 
$\chi^2$ function for $y_{t} - y_{\bar t}$ for various
$t \bar t$ polarizations.  The
solid line shows $t{(+)} {\bar t}{(+)}$ and coupling $a_{1}$.
The short-dashed line is $t{(+)} {\bar t}{(+)}$ and coupling $a_{2}$.
The dotted line shows $t{(+)} {\bar t}{(-)}$
as a function of coupling $a_{2}$.} 
\label{ydiffpolx2}
\end{figure}

\begin{figure}
\centerline{\hbox{
\psfig{figure=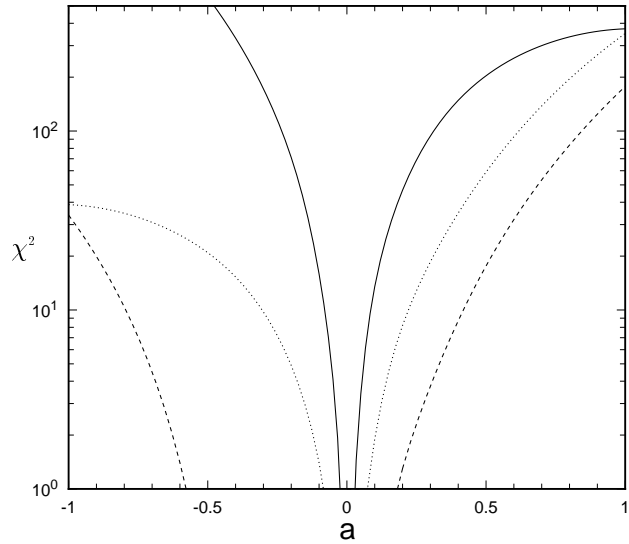,height=3.0in}}}
\caption{ 
$\chi^2$ function for $y_{t \bar t}$ for polarizations
$t{(+)} {\bar t}{(+)}$ as a function of $a_{1}$
(solid line) and $a_2$ (short dashed line) and for polarizations
$t(+) {\bar t}(-)$ as a function of $a_2$ (dotted line).
For coupling $a_{1}$ $\chi^2$ values for
$t{(+)} {\bar t}{(-)}$ are smaller than 1.0.} 
\label{yttpolx2}
\end{figure}

Finally, we combine the $t \bar t$ polarization information with
the effects upon the shape of the rapidity distributions,
$y_t - y_{\bar t}$
and $y_{t \bar t}$, for two separate choices of $t \bar t$
polarizations: $t(+) \bar t(+)$ and $t(+) \bar t(-)$,
thus combining all of the (possibly in the case of $t$ polarization)
measurable quantities considered
in this study.
The results for the $\chi^2$ functions are
presented in Fig. \ref{ydiffpolx2} and Fig. \ref{yttpolx2}
respectively.  At the 95\% C.L., we find that the constraints
on $a_1$ do not change appreciably, because the scalar operator
does not have a large effect on the shape of the rapidity
distributions, while those on $a_2$ are improved some-what by
the $y_t - y_{\bar t}$ analysis.  These results are summarized
in Table \ref{boundspol2}.

\section{Conclusions}
\label{conclu}
\indent

In this paper, we present a complete, realistic calculation of the
rate for producing $t \bar t$ pairs
at the LC, an
$e^+ e^-$ collider with $\sqrt{s} = 1.5$ TeV and
$200 \;{\rm fb}^{-1}$ of integrated luminosity.
We have considered both
$t \bar t \nu \bar \nu$ and $t \bar t \gamma$ production, within the
frame-work of the the {\it Higgsless} Standard model, an effective
theory in which the SU(2$)_L \times$ U(1$)_Y$ gauge symmetry
is realized non-linearly.
In order to parameterize the effect of ``new physics'' coming from
a high energy scale, we include the possibility of
anomalous dimension
five \wwtt local operators.  We have chosen to focus on a scalar
($O_{g {\cal W}{\cal W}}^{(5)}$) and a tensor
($O_{\sigma {\cal W}{\cal W}}^{(5)})$)
operator as representative terms in the
effective lagrangian which contribute to
scattering amplitudes at energy order $E^3$, in the S-wave
and P-wave channels respectively.  Comparing the rate for
$e^+ e^- \ra t \bar t \nu \bar{\nu}$ to previous work
\cite{top5}, in which the effective $W$ approximation was
used to estimate this rate, we find that the effective $W$
method is in good agreement with the full calculation (to
within a factor of 2).

We found that by studying the total production rate, it is possible
to constrain a dimension five anomalous scalar \wwtt coupling by
$-0.13 \leq a_1 \leq 0.18$, and a tensor \wwtt coupling by
$-0.9 \leq a_2 \leq 0.2$.  However, one can hope to
improve these bounds by considering the effect of the operators on
the rapidity distributions $y_{t \bar t}$ and $y_t - y_{\bar t}\;$ to
$-0.10 \leq a_1 \leq$ 0.12 and $-0.3 \leq a_2 \leq$ 0.2.
In addition, since the deviations from the 
{\it Higgsless} SM distributions
depend on $a_1$ and $a_2$, we can use the $\chi^2$ analysis
to decide whether the anomalous effect comes from the scalar
coupling, the tensor coupling, or from both. 

We have also studied the improvements in constraining $a_1$ and
$a_2$ resulting from a polarized electron and/or positron beam.
We find that an improvement of
about 43\% in the bounds on $a_1$ and 11\% in the bounds
on $a_2$
result when the
electron beam is 100\% polarized, and that no large
further improvement
is expected if the positron beam is also polarized.

Should it prove experimentally feasible to reconstruct the
polarization of the $t$ and $\bar t$, we can hope to further improve
these bounds.  By considering the rapidity distributions
of $t(+) \bar t(+)$ and $t(+) \bar t(-)$ separately, one can hope
to achieve the constraints $-0.08 \leq a_1 \leq 0.08$ and
$-0.20 \leq a_2 \leq 0.20$.  
Finally, because the scalar operator does not
contribute to the $t(+) \bar t(-)$ channel, one can also use
measurements of the polarized production rates to determine
directly which
operator is responsible for an observed new physics effect,
should an excess in the $t \bar t$ with missing $p_{T_{(t \bar t)}}$
rate be observed at the LC.

Thus, we conclude that the LC would provide an excellent experiment
for probing anomalous \wwtt couplings, and thus could provide key
information on the details associated with the electroweak symmetry
breaking sector, should a light Higgs boson fail to be discovered.

\medskip
\section*{ \bf  Acknowledgments }
\indent \indent

F. Larios would like to thank Conacyt and the OAS for support.
C.--P. Yuan  was supported in part by the NSF grant
No. PHY-9507683.    
Part of T. Tait's work was completed at
Argonne National Laboratory,
in the High Energy Physics division and was supported in
part by the U.S. Department of Energy, High Energy Physics
Division, under Contract W-31-109-Eng-38.

\end{document}